\documentstyle[aps,multicol,prl,epsfig]{revtex}
\title{Spin-spin interaction and spin-squeezing in an optical lattice}

\author{Anders S\o rensen \cite{anders} and Klaus M\o lmer \cite{klaus}\\
 {\small Institute of Physics and Astronomy, University of Aarhus,\\
DK-8000 \AA rhus C, Denmark}}
\begin{document}
\draft

\maketitle

\begin{abstract}
We show that by displacing two optical
lattices with respect to each other, we may
produce interactions similar to the ones describing ferro-magnetism in
condensed matter physics.  We 
also show that   particularly simple
choices of the interaction lead to spin-squeezing, which may be
used to improve the sensitivity of atomic clocks. Spin-squeezing
is generated even with  partially, and randomly, filled lattices, and 
our proposal may  be implemented with current technology. 
\end{abstract}
\widetext
\tighten
\begin{multicols}{2}

Simulation of
quantum many-body problems on a classical   
computer is difficult because the size of the Hilbert
space grows exponentially with the number of particles. As suggested by
Feynman \cite{feynman} the growth in computational requirements is only  
linear on a quantum  computer \cite{quantumcomp}, which is itself  a
quantum many-body system, and such a  computer containing only a few tens of 
quantum bits may outperform a classical computer. A quantum
computer aimed at the solution of a  quantum problem is expected to
be easier to realize in practice than a general purpose quantum computer,
because the desired solution is governed by physical interactions which are
constrained, e.g., by locality \cite{feynman,lloyd}. In essence, such a
quantum computer is a {\it quantum simulator} with the attractive feature
that the experimentalist can control and observe the dynamics more
precisely than in the physical system of interest.
In this Letter we describe how atoms in an optical lattice may be
manipulated to simulate spin-spin interactions which are used to describe
ferro-magnetism 
in condensed matter physics. We also show that with a specific choice of
interaction we may generate spin squeezed states \cite{ueda}  which may be
used to enhance spectroscopic resolution\cite{winsq}, e.g., in atomic
clocks.  

\begin{figure}
\begin{center} 
 \epsfig{file=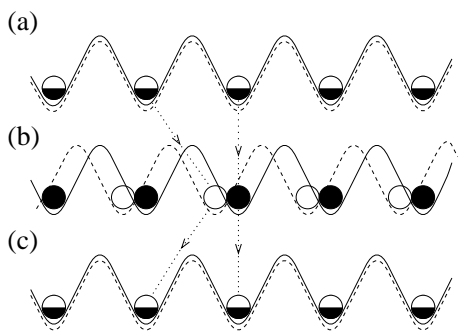,angle=270,width=6cm}
\end{center}
\vspace{0.5cm}
 \caption{(a) Two overlapping  lattices trapping the two internal states
   $|0\rangle$ (black circle) and 
   $|1\rangle$ (white circle). By resonant laser
   pulses the atoms 
   can be prepared in any superposition of the two internal states. (b) The 
   lattices are displaced  
   so that if an atom is in the $|1\rangle$ state, it
   is moved close to  the neighbouring atom if this is in
   $|0\rangle$ causing an interaction between the two atoms. (c) The
   lattices are returned to their initial position and the atoms may be
   prepared in new superpositions by resonant pulses.}
   \label{displace}
\end{figure}

In Refs. \cite{gatecirac,gatebrennen} two different methods to
perform a coherent evolution 
of the joint state of pairs of atoms  in
an optical lattice were proposed. 
Both methods
involve displacement of two identical optical lattices with respect to each
other. Each lattice traps one of the two internal states $|0\rangle$ and
$|1\rangle$ of the atoms. Initially, the atoms are in the same internal
state $|0\rangle$, the two lattices are
on top of each other and the atoms are assumed to be cooled to the
vibrational ground state in the lattice. Using a resonant pulse the atoms
may be prepared in any superposition of the two internal states. The
lattice containing the $|1\rangle$ component of the wavefunction is
now displaced so that if an  atom (at the lattice site $k$) is in
$|1\rangle$, 
it is transferred to 
the vicinity of
the neighbouring atom (at the lattice site $k+1$) if this is in
$|0\rangle$, causing an interaction between the two atoms. See
Fig. \ref{displace}.  
The procedures
described in this Letter  follow the proposal in Ref. \cite{gatecirac}
where, the atoms interact 
through controlled collisions.  Also 
the optically induced dipole-dipole interactions proposed in
\cite{gatebrennen} may be adjusted  to fit into this framework.
After the interaction,
the lattices are returned to their initial position and the internal states
of each atom may again be subject to single particle unitary evolution. The total 
effect of the displacement and the
interaction with the neighbour can be tailored to yield a certain
phaseshift $\phi$ on the 
$|1\rangle_k |0\rangle_{k+1}$ component of the wavefunction, i.e., 
\begin{eqnarray}
  |0\rangle_k|0\rangle_{k+1}\rightarrow&
  |0\rangle_k|0\rangle_{k+1} &|0\rangle_k|1\rangle_{k+1}
  \rightarrow |0\rangle_k |1\rangle_{k+1}  
  \nonumber \\ 
  |1\rangle_k|0\rangle_{k+1}\rightarrow& e^{i\phi}|1\rangle_k
  |0\rangle_{k+1}\hspace{0.25cm}&|1\rangle_k|1\rangle_{k+1} \rightarrow
  |1\rangle_k|1\rangle_{k+1}, 
  \label{phaseshift}
\end{eqnarray}
where $|a\rangle_k$ ($a=0$ or $1$) refers to the state of the atom at the
$k$'th lattice site.
In \cite{gatecirac,gatebrennen} it is suggested to build a general purpose
quantum computer in an optical lattice. Such a general computer requires
two-atom  gates, which may be
accomplished through the dynamics in (\ref{phaseshift}) and single
atom control, which is possible by directing a laser beam on each atom. We shall
show that even without allowing
access to
the individual atoms, the lattice may be used to perform a highly non-trivial
computational task: Simulation of a ferro-magnet.

Our two level quantum systems  conveniently describe  spin $1/2$
particles with the two states $|0\rangle_k$ and $|1\rangle_k$ representing
$|jm\rangle_k=|1/2, -1/2\rangle_k$ and $|1/2, 1/2\rangle_k$, where states
$|jm\rangle_k$ are eigenstates of the $j_{z,k}$-operator
$j_{z,k}|jm\rangle_k=m|jm\rangle_k$ ($\hbar=1$). The phase-shifted
component of the  wavefunction in
Eq. (\ref{phaseshift}) may be isolated by applying the operator
$(j_{z,k}+1/2)(j_{z,k+1}-1/2)$, and the total evolution composed of the
lattice translations and the interaction induced phaseshift 
may be described by the  unitary operator $e^{-iHt}$ with
Hamiltonian $H=\chi(j_{z,k}+1/2)(j_{z,k+1}-1/2)$ and time $t=\phi / \chi$. In
a filled lattice  the
evolution is  described by the Hamiltonian $H=\chi \sum_k
(j_{z,k}+1/2)(j_{z,k+1}-1/2)$, and if we are only interested in the bulk
behaviour of the atoms we may apply periodic boundary conditions, so that the
Hamiltonian reduces to
\begin{equation}
 H_{zz}= \chi \sum_{<k,l>} j_{z,k}j_{z,l},
 \label{kunz}
\end{equation}
where the sum is over nearest neighbours. By appropriately displacing the
lattice  we may extend the sum to
nearest neighbours in two and three dimensions. $H_{zz}$ coincides with the
celebrated Ising-model Hamiltonian
\cite{ising,reif} introduced to describe ferro-magnetism. Hence, by elementary
lattice displacements we perform a quantum simulation of a
ferro-magnet. 

A more general
Hamiltonian  of the type
\begin{equation}
 H_f=\sum_{<k,l>} \chi j_{z,k}j_{z,l}+ \eta j_{x,k}j_{x,l} + \lambda j_{y,k}j_{y,l}
 \label{ferro}
\end{equation}
 may be engineered using multiple resonant pulses and
displacements of the lattices: A resonant $\pi/2$-pulse acting
simultaneously on all atoms rotates the $j_z$-operators into
$j_x$-operators,  $e^{ij_{y,k}\pi/2}
j_{z,k}e^{-ij_{y,k}\pi/2}=j_{x,k}$. Hence, by applying  
$\pi/2$-pulses, in conjunction with the   displacement sequence, we turn
$H_{zz}$ 
into $H_{xx}$, the second term in Eq. (\ref{ferro}).  Similarly
we may produce $H_{yy}$, the third term in
Eq. (\ref{ferro}), and by adjusting
the duration of the 
interaction with the neighbours we may  adjust the
coefficients $\chi$, $\eta$ and $\lambda$  
to any values. We cannot, however, produce $H_f$ by simply applying
$H_{zz}$ for the desired time $t$, followed by 
$H_{xx}$ and $H_{yy}$, because the different Hamiltonians do not
commute. Instead we apply a physical implementation of a well-known
numerical scheme: The split operator technique. If we choose  short
time steps, i.e.,  small 
phaseshifts $\phi$ in Eq.  (\ref{phaseshift}), the error will only be of
order $\phi^2$, and  by
repeated application of $H_{zz}$, $H_{xx}$ and $H_{yy}$, we may
stroboscopically  approximate $H_f$. 
  
\begin{figure} 
\begin{center}
  \epsfig{file=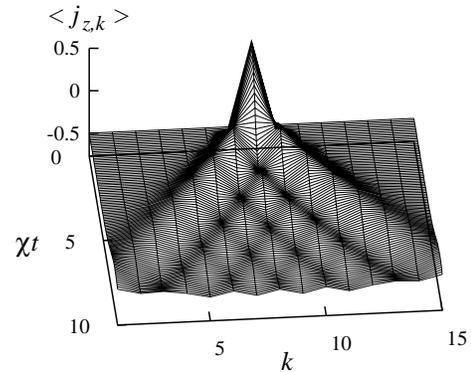,angle=270,width=6cm}
\end{center}

 \caption{Propagation of a spin wave in a one dimensional string. The
   central atom is flipped at $t=0$, and repeated application of $H_{zz}$,
   $H_{xx}$ and $H_{yy}$ results in a wave propagating to the
   left and right. The figure shows the evolution of $<j_{z,k}>$ for all
   atoms ($k$).} 
  \label{wave} 
\end{figure}

For a few atoms the system may be simulated numerically on a classical
computer. 
In Fig. \ref{wave} we show the propagation of a spin wave in a
one-dimensional string of 15 atoms which are initially in the
$|1/2,-1/2\rangle$ 
state. The central atom is flipped at $t=0$ and a spin wave propagates to
the left and right. The figure shows the
evolution of $<j_{z,k}>$ for all atoms, obtained by repeatedly
applying the Hamiltonians $H_{zz}$, $H_{xx}$ and $H_{yy}$  with
$\chi=\eta=\lambda$ and
periodic boundary conditions. Small time steps
$dt=0.1\chi^{-1}$ result in a stroboscopic approximation  almost
indistinguishable  from the
results of a direct numerical integration of $H_f$.

A host of magnetic phenomena may be simulated on our optical lattice:
Solitons,  topological excitations, two magnon bound states,
etc. By pumping a fraction of the atoms into the $|1/2,1/2\rangle$ state,
we may  also
perform micro-canonical ensemble calculations  \cite{reif} for non-vanishing
temperature. Other procedures for introducing a non-vanishing temperature are
described in Ref. \cite{lloyd}.

We now show how to generate spin squeezed states using the same techniques
as discussed above.  Signals obtained in spectroscopic
investigations of a sample of two level atoms are  expressed by
 the collective
spin operators $J_i=\sum_k j_{i,k}$, and their quantum mechanical
uncertainty limits 
the measurement accuracy, and  e.g., the performance of atomic clocks. In
standard spectroscopy with $N$ uncorrelated atoms starting in the
$|1/2,-1/2\rangle$ state,  the uncertainties $\Delta
J_x=\sqrt{\langle J_x^2\rangle-\langle J_x\rangle^2}$ and $\Delta J_y$ are
identical,   and the
standard quantum limit resulting from the uncertainty relation of angular
momentum operators
\begin{equation}
 (\Delta J_x)^2(\Delta J_y)^2\geq\left| <J_z/2>\right| ^2
 \label{heisenberg}
\end{equation}
predicts a spectroscopic sensitivity  proportional to
$1/\sqrt{N}$. Polarization rotation spectroscopy and high
precision atomic fountain clocks are
now limited by this sensitivity \cite{jens,precision}. In \cite{ueda} it is
suggested  to produce spin squeezed states which redistribute the uncertainty
unevenly between components like $J_x$ and $J_y$ in (\ref{heisenberg}), so
that measurements, sensitive to the component with reduced uncertainty,
become more precise. Spin squeezing resulting from absorption of
non-classical light has been suggested  \cite{kuzmich} and demonstrated
experimentally \cite{jan}. Ref. \cite{ueda} presents an analysis of 
squeezing obtained from
the non-linear couplings  $H=\chi J_x^2$ and
$H=\chi(J_x^2-J_y^2)$.
For neutral atoms, such a coupling  has been suggested in the
spatial overlap of two components of a Bose-Einstein condensate
\cite{becoverlap}. Spin squeezing in an  optical lattice has two main
advantages compared to 
the condensates: The interaction can be turned on and off easily, and the
localization at lattice sites increases the density and thus the
interaction strength. The product of
two collective spin operators involve terms $j_{x,k}j_{x,l}$ for all
atoms $k$ and $l$, and this coupling may be produced by
displacing the lattices 
several times so that the $|1/2,1/2\rangle$  component of  each atom visits
every 
lattice site and interacts with all other atoms. In a
large lattice such multiple displacements  are not desirable.
We shall show, however,  that substantial spin-squeezing occurs through
interaction with only a few nearby atoms,  i.e., for Hamiltonians
\begin{equation}
 H=\sum_{k,l} \chi_{k,l} j_{x,k}j_{x,l} 
 \label{x}
\end{equation}
and
\begin{equation}
 H=\sum_{k,l}  \chi_{k,l} (j_{x,k}j_{x,l}- j_{y,k}j_{y,l}),
 \label{xy}
\end{equation}
where the coupling constants $\chi_{k,l}$  between atoms $k$ and $l$
vanishes  except for a small
selection of displacements of the lattices. 

Expectation values of relevant angular momentum operators and the variance
of the spin operator 
$J_{\theta}=\cos(\theta)J_x+\sin(\theta)J_y$ may be calculated for an
initially uncorrelated state with all atoms in $|1/2,-1/2\rangle$,
propagated by the simple coupling (\ref{x}).  If each atom visits one  
neighbour $\chi_{k,l}=\chi \delta_{k+1,l}$, we get the time dependent
variance of the 
spin component $J_{-\pi/4}=\frac{1}{\sqrt{2}}(J_x-J_y)$
\begin{equation}
 (\Delta J_{-\pi/4})^2 =\frac{N}{4}\left[1+\frac{1}{4} \sin^2(\chi
 t)-\sin(\chi t)\right]. 
 \label{dj}
\end{equation}
The mean spin vector is in the negative $z$ direction and has the expectation
value 
\begin{equation}
 <J_z>=-\frac{N}{2}\cos^2(\chi t).
 \label{jz}
\end{equation}
For small values of $\chi t$, $\Delta J_{-\pi/4}$ decreases linearly with
$\chi t$ whereas $|<J_z>|$ decreases proportional to $(\chi t)^2$, hence
$\Delta J_{-\pi/4}$ 
falls below $|<J_z/2>|$, and the spin is squeezed.

\begin{figure} 
 {\center 
 \begin{minipage}{4cm}
  \epsfig{file=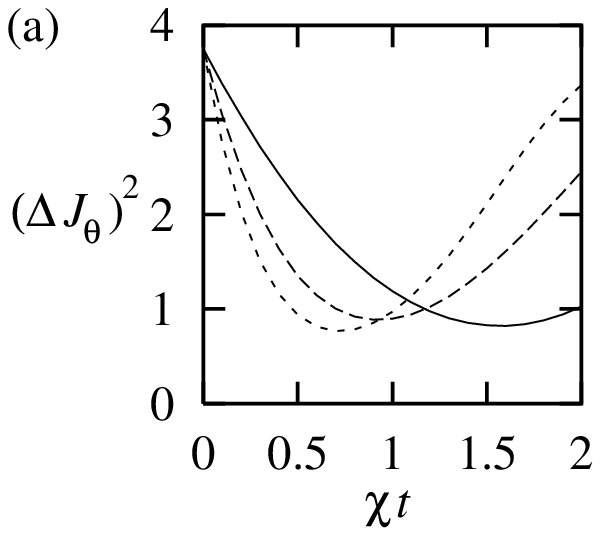,angle=270,width=4cm}
 \end{minipage}
\hspace{0.5cm}
 \begin{minipage}{4cm}
  \epsfig{file=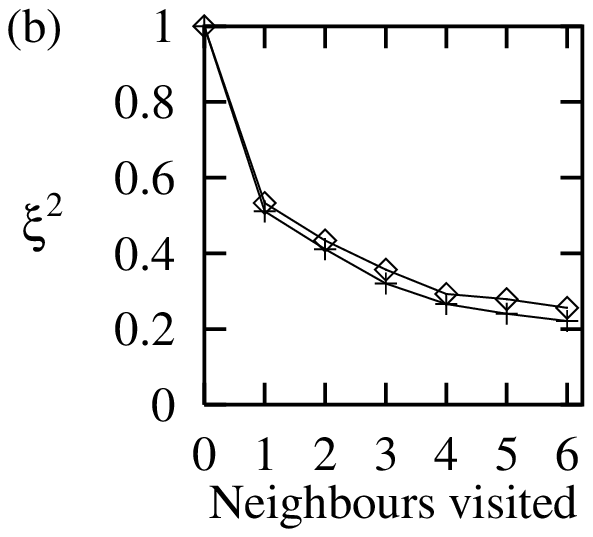,angle=270,width=4cm}
 \end{minipage}} 
 \vspace{0.5cm}
 \caption{Squeezing in a
   one-dimensional lattice with 15 atoms. (a) Evolution of $(\Delta
   J_\theta)^2$ with the coupling (\ref{x}) and interaction with 1, 2, and
   3  neighbours (full, dashed, and short dashed line, respectively). (b) The
   optimal value of the squeezing parameter $\xi^2$ obtained with the
   coupling (\ref{x}) ($\diamond$) and (\ref{xy}) ($+$). Lines are shown to
   guide to the eye.} 
  \label{fuldsqueez} 
\end{figure}

In Fig. \ref{fuldsqueez} we show numerical results for 15 atoms in a
one-dimensional lattice with periodic boundary 
conditions. Fig. \ref{fuldsqueez} (a) shows the evolution of $(\Delta
J_\theta)^2$ when we apply the
coupling (\ref{x}) and visit 1, 2, and 3 neighbours.  The
squeezing angle 
$\theta=-\pi/4$  is optimal 
for short times $\chi t << 1$. For longer times the optimal  angle deviates
from $-\pi/4$, and we plot the variance  $(\Delta J_{\theta})^2$ minimized
with 
respect to the angle $\theta$. We
assume the same phaseshift for all collisions, i.e., all non-vanishing
$\chi_{k,l}$ are identical.  

For spectroscopic investigations not only the variance of a spin component
is relevant. In \cite{winsq}
it is shown that if spectroscopy is performed with $N$
particles, the reduction in the frequency
variance due to squeezing is given by the quantity 
\begin{equation}
  \xi^2=\frac{N\langle \Delta J_\theta\rangle^2}{\langle
    J_z \rangle^2 }. 
  \label{ki}
\end{equation}
In Fig. \ref{fuldsqueez} (b) we show the minimum value of $\xi^2$ obtained
with the couplings (\ref{x}) and (\ref{xy}) as functions of the number of
neighbours visited. 
Fig. \ref{fuldsqueez} (b) shows that the coupling (\ref{xy}) produces 
better squeezing than (\ref{x}). The coupling (\ref{x}), however, is 
more attractive  from an experimental viewpoint. Firstly, all $j_{x,k}$
operators commute and we do not  have to apply several
displacements with infinitesimal durations to produce the desired
Hamiltonian.  We may simply displace 
the atoms so that they interact with  one neighbour to produce the desired
phaseshift 
$\phi$, and then go on to interact with another neighbour. Secondly, if the
$j_{x,k}j_{x,l}$  coupling involves a phaseshift $\phi$, the operator $-
j_{y,k}j_{y,l}$ requires the  opposite phaseshift $-\phi$. This
requires a long 
interaction producing $2\pi-\phi$, or a change of the interaction among the
atoms,  i.e., a change 
of the sign  of the scattering length in the implementation of
\cite{gatecirac}.

Like the analytic expression for $\chi^2$ obtained from
(\ref{dj},\ref{jz}), the results shown in 
Fig. \ref{fuldsqueez} (b) are independent of the total number of atoms as
long as this exceeds the ``number of neighbours visited''. When all lattice
sites are visited we approach the results obtained in Ref.
\cite{ueda}, i.e., a variance scaling as $N^{1/3}$ and a constant for the
couplings (\ref{x}) and (\ref{xy}).

So far we have assumed that the lattice contains one atom at each
lattice site and that all atoms  are cooled to the vibrational ground
state. The present
experimental status is that atoms can be cooled to the vibrational
ground-state, but with a filling factor below unity
\cite{groundstate}. A mean filling factor of unity is reported in
\cite{depue}, but when at most a single atom is permitted at each 
lattice site  a mean occupation of 0.44 is achieved. It has been suggested
that a 
single atom per lattice site may be achieved by filling
the lattice from a Bose-Einstein condensate \cite{filled}.  

To describe a partially filled lattice it is convenient to introduce
stochastic variables $h_k$, describing whether the $k$'th lattice site is
filled $h_k=1$ or empty $h_k=0$. 
The interaction  may be described by the Hamiltonian
$H=\sum_{k,l} \chi_{k,l} h_k(j_{z,k}+1/2)h_l(j_{z,l}-1/2)$, where the sum
is over all lattice sites $k$ and $l$. If we, rather than
just displacing the atoms in one direction, also displace the lattice in
the opposite direction, so that $\chi_{k,l}$ is symmetric in $k$ and $l$,
we may produce
the Hamiltonian
\begin{equation}
 H=\sum_{k,l} \chi_{k,l} h_k j_{x,k} h_l j_{x,l}.
 \label{partialh}
\end{equation}
This Hamiltonian models ferro-magnetism in random structures,
and it might shed light on morphology properties, and, e.g., percolation
\cite{perculation}. Here we shall restrict our analysis to spin-squeezing
aspects, since these are both of practical interest, and they represent an
ideal experimental signature of the microscopic
interaction. 

\begin{figure} 
 {\center 
 \begin{minipage}{4cm}
  \epsfig{file=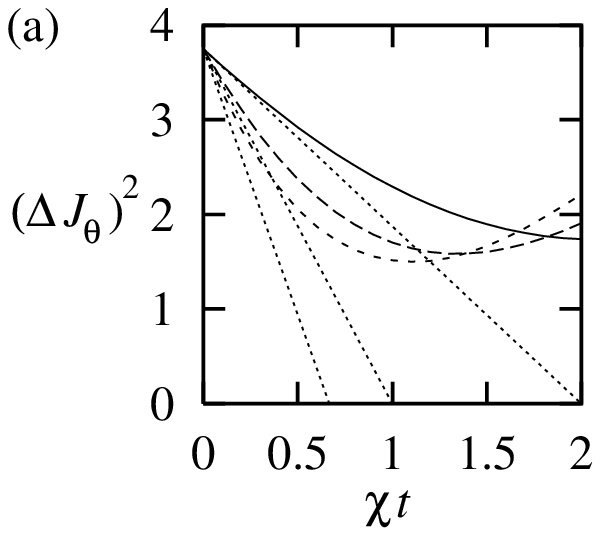,angle=270,width=4cm}
 \end{minipage}
\hspace{0.5cm}
 \begin{minipage}{4cm}
  \epsfig{file=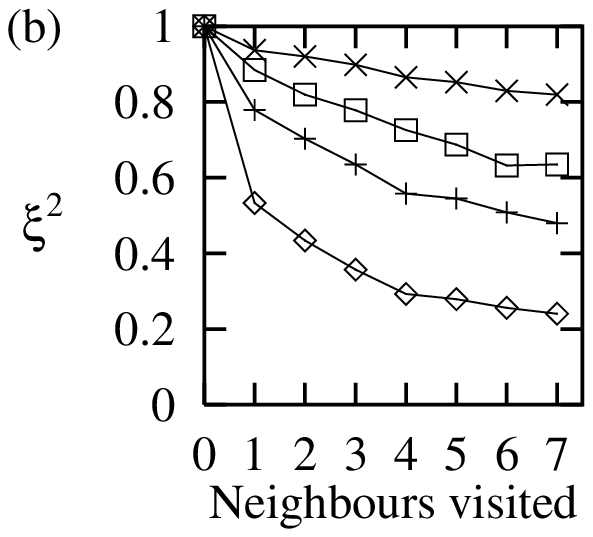,angle=270,width=4cm}
 \end{minipage}} 
 \vspace{0.5cm}
 \caption{Spin squeezing in a partially filled one dimensional lattice
   containing 15 atoms. (a) Evolution of $(\Delta J_\theta)^2$ in a lattice
   with a filling factor $p=50$\% and displacements to 1, 2, and 3
   neighbouring sites (full, dashed, and short dashed curve
   respectively). Dotted lines represent the predictions from
   Eq. (\ref{deriv}). (b)
   Minimum attainable squeezing parameter $\xi^2$ for filling factors
   $p$=100\% ($\diamond$), 50\% (+), 25\% ($\Box$), and 10\% ($\times$) as
   functions of the number of sites 
   visited.  } \label{partsqueez} 
\end{figure}

In Fig. \ref{partsqueez} we show the result of a simulation of 
squeezing in a partially filled one dimensional lattice. Each lattice site
contains  an atom with a probability $p$, and the size
of the lattice is adjusted so that it contains 15 atoms. In Fig.
\ref{partsqueez} (a) we  show the decrease in the variance of $J_\theta$,
averaged over 20 realizations and
minimized with respect to  $\theta$. Lines indicate the predictions from 
the time derivatives at $t=0$ 
\begin{eqnarray}
 \frac{d}{d t} (\Delta J_{-\pi/4})^2&=& -\frac{1}{2} \sum_{k,l}
 \chi_{k,l}  <h_k h_l> \nonumber \\
 \frac{d}{dt} J_z&=&0,                           
 \label{deriv}
\end{eqnarray}
where $<h_k h_l>$ denotes the ensemble average over the distribution of
atoms 
in the lattice, i.e., the two atom correlation function. In
Fig. \ref{partsqueez}  (b) we show the minimum value
of $\xi^2$ for different filling factors $p$ as a function of the number
of neighbours visited. The calculations confirm that even in dilute
lattices, considerable squeezing
may be achieved by visiting a few neighbours.
 
In conclusion we have suggested a method to simulate condensed matter
physics in an optical lattice, and we have shown how the dynamics may be
employed to produce spin-squeezing. We emphasize the
moderate experimental requirement for our scheme. With the two internal
states represented as hyperfine structure states in alkaline atoms, all
spin rotations may be performed by Raman or RF-pulses acting on all
atoms simultaneously, and lattice displacements may be performed by simply
rotating the polarisation of the lasers \cite{gatebrennen}. With the
parameters in \cite{gatecirac}, the duration of the sequence in
Fig. \ref{displace} can be as low as a few micro-seconds. Following our
suggestion spin-squeezing  may be produced in dilute optical lattices,
and implementation is possible with current technology. The
resulting macroscopic decrease in  projection  noise has several promising
applications in technology and quantum physics, and it provides an
experimental signature of the microscopic interaction between the atoms.

\end{multicols}
\end{document}